\documentclass[prb,twocolumn,showpacs,letterpaper,showpacs,superscriptaddress]{revtex4-1}

\usepackage{graphicx,amsmath,amssymb,amsfonts,latexsym,color,dcolumn,bm,epsfig,subfigure}
\usepackage[plainpages=false,hyperfootnotes=false,colorlinks=false]{hyperref}
\usepackage[normalem]{ulem}
\usepackage{dsfont}

\usepackage{float} 

\newcommand{\ket}[1]{\vert #1 \rangle}
\newcommand{\bra}[1]{\langle #1\vert}

\newcommand{\mean}[1]{\left\langle #1\right\rangle}
\newcommand{\abs}[1]{\left| #1\right|}
\newcommand{\Det}[1]{\left\vert #1\right\vert}

\newcommand{\Real}[1]{\mathbb{R}\text{e}\left[ #1 \right]}
\newcommand{\Imag}[1]{\mathbb{I}\text{m}\left[ #1 \right]}

\newcommand{\Comment}[1]{}

\begin{document}

\title{Nonlocal Optical Response in Topological Phase Transitions in the Graphene Family}

\author{Pablo Rodriguez-Lopez}
\affiliation{Materials Science Factory, Instituto de Ciencia de Materiales de Madrid, ICMM-CSIC, Cantoblanco, E-
28049 Madrid, Spain}
\affiliation{Department of Physics, University of South Florida, Tampa FL, 33620, USA}
\affiliation{GISC-Grupo Interdisciplinar de Sistemas Complejos, 28040 Madrid, Spain}

\author{Wilton J. M. Kort-Kamp}
\affiliation{Center for Nonlinear Studies, MS B258, Los Alamos National Laboratory, Los Alamos, NM 87545, USA}
\affiliation {Theoretical Division, MS B213, Los Alamos National Laboratory, Los Alamos, NM 87545, USA}

\author{Diego A. R. Dalvit}
\affiliation {Theoretical Division, MS B213, Los Alamos National Laboratory, Los Alamos, NM 87545, USA}

\author{Lilia M. Woods} 
\affiliation{Department of Physics, University of South Florida, Tampa FL, 33620, USA}

\date{\today}

\begin{abstract}
We investigate the electromagnetic response of staggered two-dimensional materials of the graphene family,  including silicene, germanene, and stanene, as they are driven through various topological phase transitions using external fields. Utilizing Kubo formalism, we compute their optical conductivity tensor taking into account the frequency and wave vector of the electromagnetic excitations, and study its behavior over the full electronic phase diagram of the materials. We also consider the plasmon excitations in the graphene family and find that nonlocality in the optical response can affect the plasmon dispersion spectra of the various phases. The expressions for the conductivity components are valid for the entire graphene family and can be readily used by others.

\end{abstract}

\maketitle

\section{Introduction}
The discovery of layered materials with a honeycomb lattice has created unprecedented opportunities to study Dirac-like physics and related phenomena \cite{Neto2009,Balendhran2015}. Recent efforts in silicene, germanene, and stanene, the 2D hexagonal allotropes of Si, Ge, and Sn, respectively, have brought forward new directions in electronic, optical, and transport properties beyond graphene \cite{Castellanos-Gomez2016,Mannix2017}. 
The experimental realization of these graphene-like atom-thin layers is still in its infancy. After the initial synthesis of silicene on Ag and ZrB$_2$ substrates\cite{Vogt2012,Lin2012,Fleurence2012}, the fabrication of silicene-based field effect transistors has been reported \cite{Lelay20115}. Germanene and stanene have also been synthesized on several metallic substrates, including gold, aluminum, and Bi$_2$Te$_3$ \cite{Davila2014,Derivaz2015,Zhu2015,Davila2016}. While carbon atoms in graphene are in a planar configuration with $sp^2$ orbital hybridization, silicene, germanene, and stanene are characterized by a buckled structure with $sp^3$ orbital hybridization \cite{Cahangirov2009,Matthes2013}. 
In contrast to graphene, that has negligible spin orbit coupling (SOC), these materials have significant SOC. 
The interaction of these 2D honeycomb systems with a circularly polarized laser and/or an electrostatic field perpendicular to the surface can be used to achieve several electronic phases \cite{Lin2011,Ezawa2012,Tabert2013,Tabert2013a,Houssa2016}. The interplay between Dirac-like physics, SOC, and external fields also drives associated topological phase transitions in the Casimir interaction in the graphene family  \cite{Rodriguez-Lopez2017}. Furthermore, proximity with an $s$-wave superconductor and photo-irradiation may also transform a staggered layer into a topological superconductor with Majorana fermions \cite{Ezawa2015}. All these fascinating properties offer new prospects for future applications in areas such as electronics, spintronics, and valleytronics. 

The dynamical conductivity tensor {\boldmath$\sigma$}, which is directly related to the underlying electronic structure, is a basic quantity to understand light-matter interactions in these materials. For instance, the Dirac-like energy spectrum taken into {\boldmath$\sigma$} leads to the well-known $\pi \alpha$ infrared absorption, where  $\alpha$ is the fine structure constant \cite{Matthes2013,Nair2008,Mak2008}. Modeling of several optoelectronic devices, e.g. tunable mid-IR and terahertz structures based on graphene plasmonics \cite{Low2014,deAbajo2014,Yao2013}, rely on the proper characterization of the dynamical conductivity. Several theoretical works have analyzed the optical response of graphene through its 
spatio-temporal dispersive conductivity tensor, computed either via the Kubo formalism \cite{Falkovsky2007}, the semi-classical Boltzmann transport equation \cite{Lovat2013}, and through the polarization tensor \cite{Wunsch2006,Hwang2007}.
The nonlocal optical response of the other members of the graphene family has been recently studied using the polarization tensor approach \cite{Tabert2014,Wu2016}, but these works are limited to phases with trivial topology when no external polarized laser is applied.

In this paper, we present a unified description of the nonlocal  dynamical conductivity tensor of the graphene family materials based on the Kubo formalism, including non-trivial topological phases that arise from an externally applied circularly polarized laser and an electrostatic field. The derived analytical formulas give explicit expressions for the conductivity components, in which  the inter and intraband contributions, the chemical potential, and carrier relaxation are also taken into account. Our results show that the frequency and wave vector dependences lead to a rich structure of the optical response, in particular various anisotropic behaviors of the conductivity tensor in the phase space diagram. 
We also investigate plasmon excitations in the graphene family materials, and find that nonlocality  affects their dispersion spectra along different topological phases, especially at large wave vectors. 


\section{Theoretical Background}
\begin{figure}
\centering
\includegraphics[width=\linewidth]{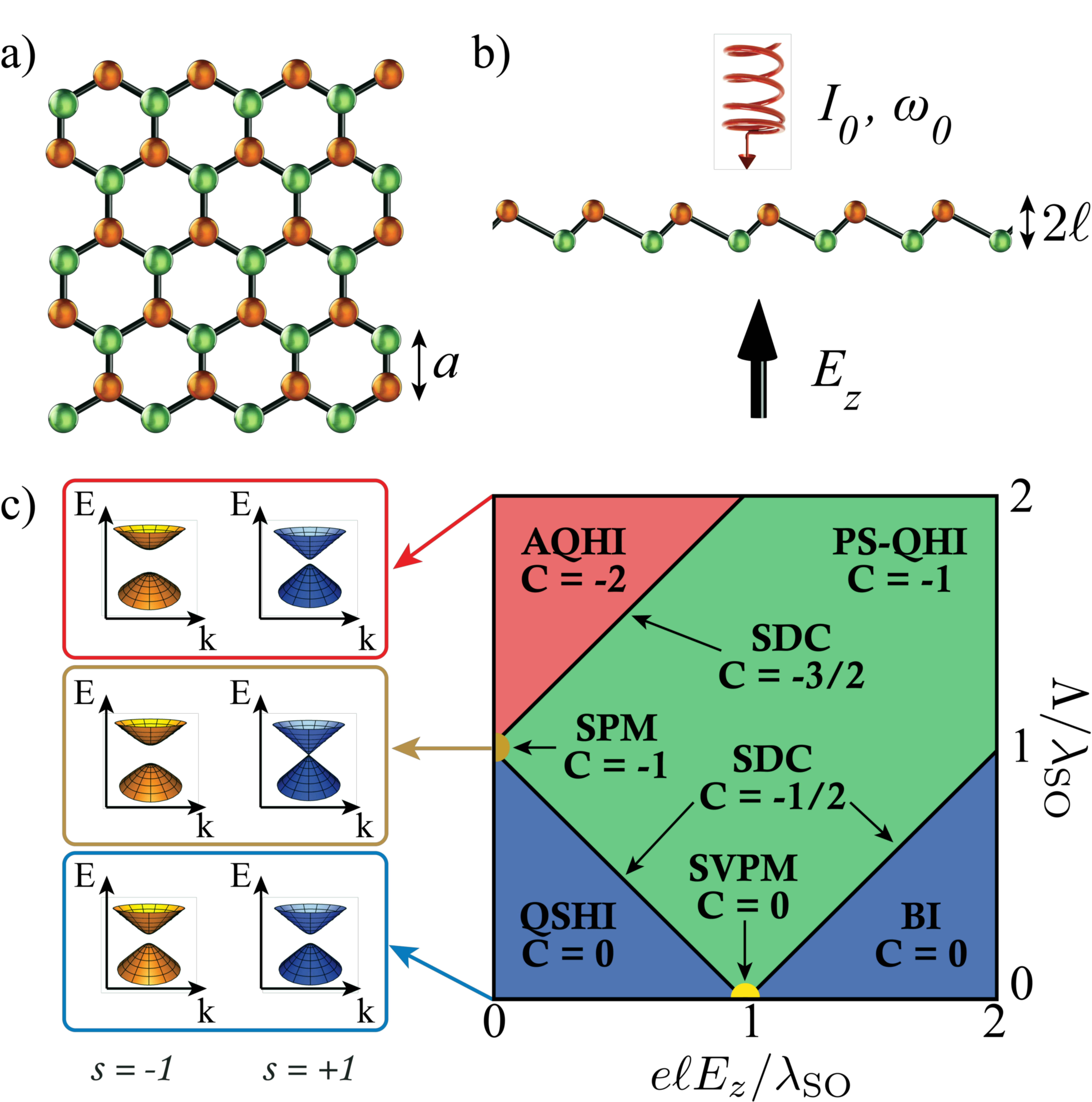}
\caption{(Color online) (a) Top view of a honeycomb layer from the graphene family. The two inequivalent atoms are shown in green and brown colors and the nearest neighbor distance is $a$. (b) Side view of a staggered layer. The lattice buckling is denoted as $2\ell$. The impinging circularly polarized light with intensity $I_0$ and frequency $\omega_0$ and external static electric field with strength $E_z$ applied perpendicular to the surface are also shown. (c) Schematics of the Dirac-like spectra for some of the electronic Hall phases of the graphene family. The first quadrant of the phase diagram of the materials in the $(e\ell E_z , \Lambda)$ plane in units of $\lambda_{SO}$ is also given. The  Chern number $C$ denotes the different electronic phases with acronyms defined in the text.}
\label{Phasediagram}
\end{figure}

Graphene, silicene, germanene, and stanene have a layered hexagonal lattice (Fig. \ref{Phasediagram}). The low energy band structure can be obtained from a nearest neighbor tight binding model. The resulting Dirac-like Hamiltonian is valid for the entire graphene family, which also includes an external static electric field $E_{z}$ perpendicular to the layer and an  applied circularly polarized laser characterized by 
$\Lambda = \pm 8 \pi \alpha v_{F}^2 I_{0}/\omega_{0}^3$, where $v_F$ is the Fermi velocity,
$I_0$ is the intensity of the laser, $\omega_0$ is its frequency, and $\pm$ denotes the left and right polarizations \cite{Lin2011,Ezawa2012,Ezawa2013}
\begin{eqnarray}\label{Eq_Hamiltonian}
H^{\eta}_s = \hbar v_{F}\left[\eta k_{x}\tau_{x} + k_{y}\tau_{y} \right] + \tau_{z}\Delta^{\eta}_s, 
\end{eqnarray}
\begin{eqnarray}\label{Mass-gap}
\Delta^{\eta}_s = \eta s\lambda_{SO} - e \ell E_{z} - \eta\Lambda.
\end{eqnarray} 
The given Hamiltonian $H^{\eta}_s$ and mass gap $\Delta^{\eta}_s$ are per Dirac cone characterized by the valley index $\eta = \pm 1$ and spin index $s=\pm 1$. Also, $\tau_{i}$ are the Pauli matrices, the in-plane components of the 2D wave vector are $k_{x,y}$, $\lambda_{SO}$ is the spin-orbit coupling, and $2\ell$ is the distance between the sub-lattices (see Fig. \ref{Phasediagram}). Using Eqs. \ref{Eq_Hamiltonian}, \ref{Mass-gap}, the energy dispersion is obtained as $\epsilon_{\textbf{k}}^{\lambda} = \lambda\sqrt{ \hbar^{2}v_{F}^{2}k^{2} + (\Delta^{\eta}_s)^{2} }$, where $\lambda=\pm$ represents the upper ($+$) and lower ($-$) bands of the 2D Dirac cone and $k^2=k_x^2+k_y^2$. The parameters for staggering, SOC strength, Fermi velocity, and other structure characteristics are available for the entire graphene family \cite{Ezawa2015b}.
To simplify notation, in the following we will not explicitly write the valley and spin indexes in the expressions for the conductivity tensors and the gaps. Whenever necessary, we will emphasize when a summation over these indexes is needed. 

The external factors can be used to control the mass gaps for the different Dirac cones. As a result, it is possible to obtain a rich variety of electronic phases associated with the Hall effect.  Depending on how many Dirac cones have non-zero mass gaps, quantum states such as Quantum Spin Hall Insulator (QSHI), Spin Polarized Metal (SPM), Spin Valley Polarized Metal (SVPM), Anomalous Quantum Hall Insulator (AQHI), Band Insulator (BI),  Single Dirac Cone (SDC), and Polarized Spin Quantum Hall Insulator (PS-QHI) become possible. Each state can be characterized with a Chern number 
$C = {\sum}'_{s,\eta= \pm 1}\eta\,\text{sign}\left[\Delta^{\eta}_s \right]$, where the prime in the summation indicates that only terms with $\Delta^{\eta}_s \neq 0$ should be included \cite{Ezawa2013,Ezawa2015b,Rodriguez-Lopez2017}. Fig. \ref{Phasediagram}(c) shows the first quadrant of the complete phase diagram for these materials. 

In this work we utilize the standard Kubo formalism \cite{Kubo1,Kubo2} to calculate the dynamical optical response for the graphene family by taking into account its dependence on frequency $\omega$ and wave vector $\textbf{q}=(q_x,q_y)$ of the photons. The components of the conductivity tensor per Dirac cone are given as 
\begin{widetext}
\begin{eqnarray}\label{General_Kubo_Formula}
\sigma_{ij}(\omega,\textbf{q}) & = & - i 4 \sigma_{0}\hbar^{2} \sum_{\lambda,\lambda'}\int \frac{d^{2}\textbf{k}}{(2\pi)^{2}}
\frac{\bra{u_{\textbf{k}}^{\lambda}} v_{i} \ket{u_{\textbf{k}+\textbf{q}}^{\lambda'}} \bra{u_{\textbf{k}+\textbf{q}}^{\lambda'}} v_{j} \ket{u_{\textbf{k}}^{\lambda}} }{\hbar(\omega + i\Gamma) + \epsilon_{\textbf{k}}^{\lambda} - \epsilon_{\textbf{k}+\textbf{q}}^{\lambda'} }
\frac{ n_{F}(\epsilon_{\textbf{k}}^{\lambda}) - n_{F}(\epsilon_{\textbf{k}+\textbf{q}}^{\lambda'}) }{\epsilon_{\textbf{k}}^{\lambda} - \epsilon_{\textbf{k}+\textbf{q}}^{\lambda'} },
\end{eqnarray}
\end{widetext}
where $\sigma_0=\alpha c/4$ is the universal conductivity of graphene ($\alpha = \frac{e^{2}}{\hbar c}$ is the fine structure constant),
$\textbf{v}  = \nabla_{\bm{k}}H_{s}^{\eta}/\hbar = v_{F}(\eta\tau_{x}, \tau_{y})$ is the velocity operator, and $u_{\textbf{k}}^{\lambda} = \left(-(\Delta+\epsilon_{\textbf{k}}^{\lambda})\frac{k_{y} + i\eta k_{x}}{k}, \hbar v_{F}k \right)/\left( \sqrt{2\epsilon_{\textbf{k}}^{+}(\epsilon_{\textbf{k}}^{+} + \lambda\Delta) } \right)$ are the eigenstates of the Hamiltonian with corresponding eigenenergies $\epsilon_{\textbf{k}}^{\lambda}$ \cite{Ezawa2015b}. The Fermi-Dirac distribution function is given by $n_F(\epsilon_\textbf{k}^{\lambda})=(e^{(\epsilon_\textbf{k}^{\lambda}-\mu)/k_B T}+1)^{-1}$ and the explicit analytical expressions for the optical response presented below are for the case when the temperature is $T=0$ K. 


\section{Results and Discussions} 

The components of the conductivity tensor can be conveniently given by separating between longitudinal $\sigma_L$, transverse $\sigma_T$, and Hall $\sigma_{H}$, contributions\cite{Zeitlin1995,Wooten1972}
\begin{eqnarray}
\label{General_non_local_conductivity_form}
\sigma_{ij}(\omega, \textbf{q})&=& \frac{q_{i}q_{j}}{q^{2}}\sigma_{L}(\omega, q) + 
\left( \delta_{ij} - \frac{q_{i}q_{j}}{q^{2}} \right)\sigma_{T}(\omega, q) \nonumber\\
&+& \epsilon_{ij}\sigma_{H}(\omega, q).
\end{eqnarray}
Here $\delta_{ij}$ is the Kronecker delta function and $\epsilon_{ij}$ is the 2D Levi-Civita symbol. 

The analytical formulas for the different types of response can be separated into two terms,
one independent of the chemical potential $\mu$ and another term for which the chemical potential is accounted for. Namely,
$\sigma_{p}(\omega, q) =  \sigma_{p,0}(\omega, q) + \Theta(\abs{\mu} - \abs{\Delta})\sigma_{p,1}(\omega, q)$, where $\Theta$ is the Heaviside step function and
$p=\{L, T, H\}$. The procedure for obtaining Eq.(\ref{General_non_local_conductivity_form}), together with details of calculating the different components of the optical conductivity, are given in the Appendix.  The obtained results are
\begin{widetext}
\begin{eqnarray}
\label{Sigma_L_0}
\sigma_{L,0}(\omega, q) & = &  \frac{\sigma_0}{4 \pi}
\frac{ - i\Omega}{\left( Q^{2} - \Omega^{2} \right)}\left[ 4\abs{\Delta} + \frac{ Q^{2} - \Omega^{2} - 4\Delta^{2}}{\sqrt{ Q^{2} - \Omega^{2} }}i\log\left( \frac{ \Omega^{2} - Q^{2} + 4\Delta^{2} - 4i\abs{\Delta}\sqrt{ Q^{2} - \Omega^{2} } }{ Q^{2} - \Omega^{2} + 4\Delta^{2} }\right)\right], \\
\label{Sigma_L_mu}
\sigma_{L,1}(\omega, q) & = &  \frac{\sigma_0}{4 \pi}
\frac{-i}{\Omega Q^{2}} 
\left[
8\Omega^{2}(\abs{\mu} - \abs{\Delta} ) + \frac{\Omega^{2}}{\sqrt{ Q^{2} - \Omega^{2} }}\mathcal{F}_{1} + \frac{Q^{2}}{\sqrt{ Q^{2} - \Omega^{2} }}\left( \Omega^{2} + 4\Delta^{2}\left( 1 - \frac{Q^{2}}{ Q^{2} - \Omega^{2} }\right)\right)\mathcal{F}_{2} \right. \nonumber \\
&& ~~~~~~~~~~~~~ \left. + 2Q \Theta\left(Q^{2} - 4\left( \mu^{2} - \Delta^{2} \right)\right)\mathcal{F}_{3}
\right],\\
\label{Sigma_T_0}
\sigma_{T,0}(\omega, q) & = & - \frac{i \sigma_0}{4\pi\Omega} 
 \!\!\left[ \frac{2\left( Q^{2} - 4\Delta^{2} \right)}{Q}\tan^{-1}\!\!\left(\frac{Q}{2\abs{\Delta}}\right) - \frac{ Q^{2} - \Omega^{2} - 4\Delta^{2} }{\sqrt{ Q^{2} - \Omega^{2} }}i \log\!\!\left(\frac{ \Omega^{2} - Q^{2} + 4\Delta^{2} - 4i\abs{\Delta}\sqrt{ Q^{2} - \Omega^{2} } }{ Q^{2} - \Omega^{2} + 4\Delta^{2} }\right)\!\!\right]\!\!,\\
\label{Sigma_T_mu}
\sigma_{T,1}(\omega, q) & = & \frac{i 2 \sigma_0}{\pi\Omega} 
\!\!\left[
\frac{\Omega^{2}}{Q^{2}}( \abs{\mu} - \abs{\Delta} ) + 
\frac{1}{8\sqrt{ Q^{2} - \Omega^{2} }}\left( \left( \frac{\Omega^{2}}{Q^{2}} - 1 \right)\mathcal{F}_{1} + \left( Q^{2} - \Omega^{2} - 4\Delta^{2}\right)\mathcal{F}_{2} \right) \right. \nonumber \\
&& \left. ~~~~~~~~ + \frac{1}{8\sqrt{ Q^{2} - \Omega^{2} }}\left( \frac{\abs{\Delta}}{2} + \frac{\mathcal{F}_{4}}{Q}\Theta\left( Q^{2} - 4\left( \mu^{2} - \Delta^{2} \right) \right) + \frac{\mathcal{F}_{5}}{Q}\Theta\left( 4\left( \mu^{2} - \Delta^{2} \right) - Q^{2} \right)\right)
\right],\\
\label{Sigma_D_0}
\sigma_{H,0}(\omega,q)   & = &   \frac{2 \sigma_0}{\pi}\frac{\eta\Delta}{\sqrt{ Q^{2} - \Omega^{2} }}\tan^{-1}\left(\frac{\sqrt{Q^{2} - \Omega^{2}}}{2\abs{\Delta}}\right), \\
\sigma_{H,1}(\omega,q) & = & - \frac{\sigma_0}{\pi}\frac{\eta\Delta}{\sqrt{ Q^{2} - \Omega^{2} }} 
\left[
\tan^{-1}\left( \frac{\Omega - 2\abs{\Delta}}{\sqrt{ Q^{2}R^{2} - (\Omega - 2\abs{\Delta})^{2} }} \right) 
- \tan^{-1}\left( \frac{\Omega - 2\abs{\mu}}{\sqrt{ Q^{2}R^{2} - (\Omega - 2\abs{\mu})^{2} }} \right) \right. \cr
&&
\left.
~~~~~~~~~~~~~~~~~~~~~~~~~~~~~~~ + 
i \log\left( \frac{\Omega + 2\abs{\mu}    + \sqrt{ ( \Omega + 2\abs{\mu}    )^{2} - R^{2}Q^{2} } }{\Omega + 2\abs{\Delta} + \sqrt{ ( \Omega + 2\abs{\Delta} )^{2} - R^{2}Q^{2} } } \right) \right],
\label{Sigma_D_mu}
\end{eqnarray}
\end{widetext}
where $\Omega  = \hbar\omega + i\hbar\Gamma$ ($\Gamma=\tau^{-1}$ accounts for the relaxation time), $Q=  \hbar v_{F} q$, and $R=\sqrt{ 1 + 4\Delta^{2}/(Q^{2} - \Omega^{2})}$. The auxiliary functions $\mathcal{F}_{1}, ... , \mathcal{F}_{5}$ depend on $\Omega$, $\mu$, $|\Delta|$, $Q$, and $R$, and are given in the Appendix.  
It is important to note that only the modulus of the mass gaps enter into the expressions for $\sigma_L$ and $\sigma_T$, while $\sigma_{H}$ has an additional dependency on the sign of the gaps through the combination
$\eta \Delta^{\eta}_s$. In the rest of this section we will focus on the real parts of the conductivity components and study their behavior in the $(\omega, q)$ plane for various points in phase space. We will refrain from showing the corresponding imaginary parts, as they have structures that can be interpreted in a similar fashion as their real counterparts. 


\begin{figure*}
\centering
\includegraphics[width=\linewidth]{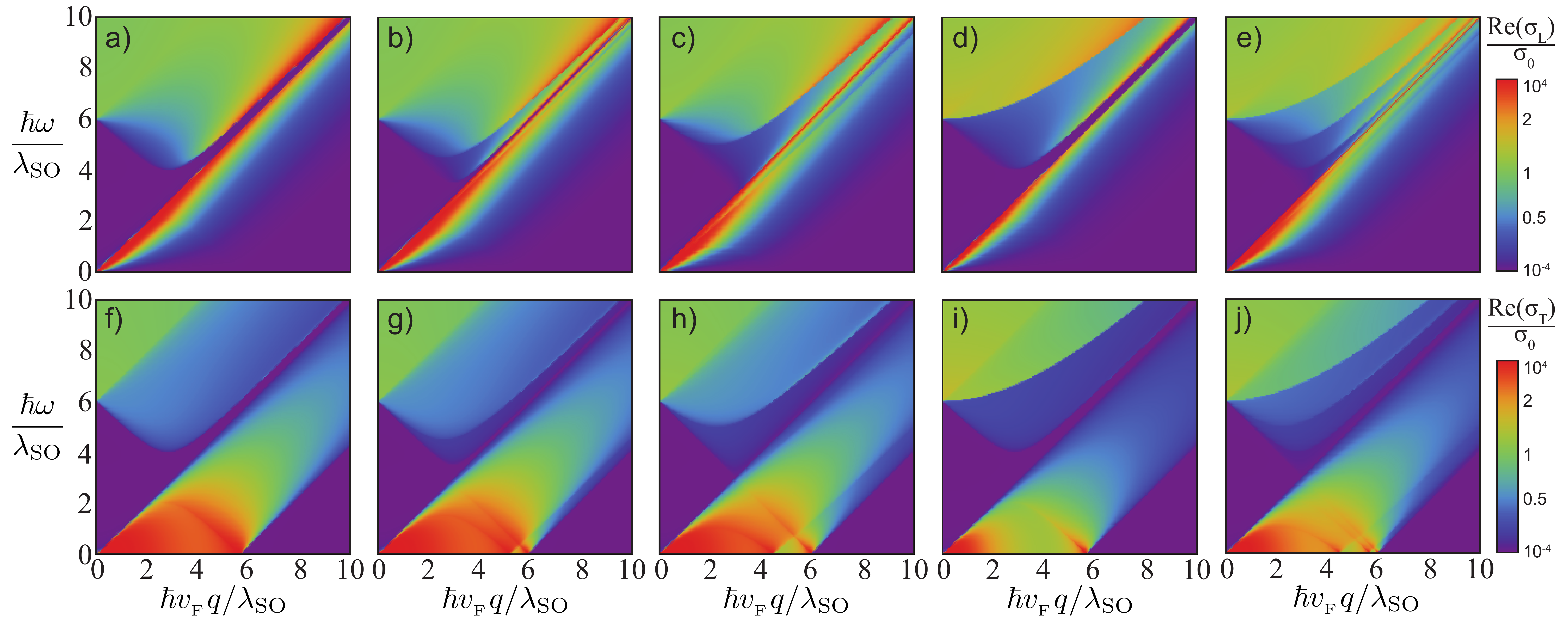}
\caption{(Color online) 
Density plots of the real part of the longitudinal $\sigma_L$ (top row) and transverse $\sigma_T$ (bottom row) dynamical conductivities in the 
$(\hbar\omega/\lambda_{SO}$, $\hbar v_F q/\lambda_{SO})$ plane for $(e \ell E_z/\lambda_{SO},\Lambda/\lambda_{SO})$ equal to $(0,0)$ (a,f), $(0,0.5)$ (b,g), $(0,1)$ (c,h), $(0,2)$ (d,i), and $(0.5,1.5)$ (e,j). Parameters are
$\mu/\lambda_{SO}=3$, and $\hbar\Gamma/\lambda_{SO}=10^{-3}$. For all conductivities the sum over valley and spin indexes has been performed.}
\label{Fig2}
\end{figure*}


In Fig. 2 we show the real part of the longitudinal $\sigma_L$ and the transverse $\sigma_T$ dynamical conductivities
in the $(\omega,q)$ plane after summation over valley and spin indexes for selected points in the electronic phase diagram of the graphene family materials. 
In the absence of electric field and laser (panels a and f), all gaps are degenerate (given by $\lambda_{SO}$) and 
the resulting conductivities are identical to those of gapped graphene (a detailed discussion of the various features appearing for gapped graphene can be found in \cite{Pyatkovskiy2009}). Let us now consider the case of applied polarized laser and no electrostatic field. For $\Lambda<\lambda_{SO}$ (QSHI phase), the degeneracy between gaps for spin up and down is broken, and the resulting density plots correspond to the weighted addition of conductivities of two gapped graphenes (panels b, g). In this  $\Lambda<\lambda_{SO}$ regime, the region $\omega> v_F q$ splits into two, with the region associated with the spin up (whose gap decreases as $\Lambda$ grows) moving toward the $\omega= v_F q$ diagonal, and the one corresponding to spin down (whose gap increases) moving away from the diagonal. For $\omega< v_F q$, the rounded red feature in the transverse conductivity splits into two contributions due to a change in the relative magnitude of the chemical potential with respect to the gap, resulting in one contribution moving to the left (spin down) and the other one moving to the right (spin up). At the phase transition point $\Lambda=\lambda_{SO}$ (SPM phase, panels c, h), the gap for spin up closes and the conductivities correspond to the weighted superposition of those of gapped and ungapped graphene, with a large increase of the optical response along $\omega= v_F q$ diagonal. 
For $\Lambda>\lambda_{SO}$, all gaps are opened and increase as $\Lambda$ grows. As a result, all the features previously described move in the same fashion (panels d, i). Finally, we show the longitudinal and transverse conductivities for a point on the SDC phase (panels e, j), in which one gap is closed and all other three have different non-zero magnitude. This allows the identification of the contribution of individual Dirac cones to the nonlocal conductivity tensor.  One should further note that while the transitions between the differently colored regions in all panels are abrupt for weakly dissipative materials, the corresponding boundaries become smoother for higher losses \cite{Pyatkovskiy2009}. 
We also note that both the longitudinal and transverse conductivities have a reflection symmetry with respect to the $\Lambda=e \ell E_z$ diagonal in the electronic phase diagram of Fig. 1. Indeed, by using that  $\Delta^{\pm 1}_s(e \ell E_z,\Lambda)=\pm \Delta^{\pm 1}_{\pm s}(\Lambda,e \ell E_z)$ and that $\sigma_L$ and $\sigma_T$ depend on the modulus of the gaps $\Delta^{\eta}_s$, it is easy to prove that $\sum_{\eta,s} \sigma_{L/T}(\omega,q; e \ell E_z,\Lambda) = 
\sum_{\eta,s} \sigma_{L/T}(\omega,q;\Lambda,e \ell E_z)$. An important consequence of the symmetry relation is that these conductivities cannot contain any information about topology. Furthermore, 
it allows us to find the longitudinal and transverse conductivities in the full phase diagram by calculating them in just half of it. 
For example, the density plots at points $(e\ell E_z/\lambda_{SO}, \Lambda/\lambda_{SO})$ equal to $(0.5,0)$,  $(1,0)$,  $(2,0)$, and  $(1.5,0.5)$ correspond to those shown in panels (b, g), (c, h), (d, i), and (e, j) of Fig. 2, respectively.

\begin{figure*}
\centering
\includegraphics[width=\linewidth]{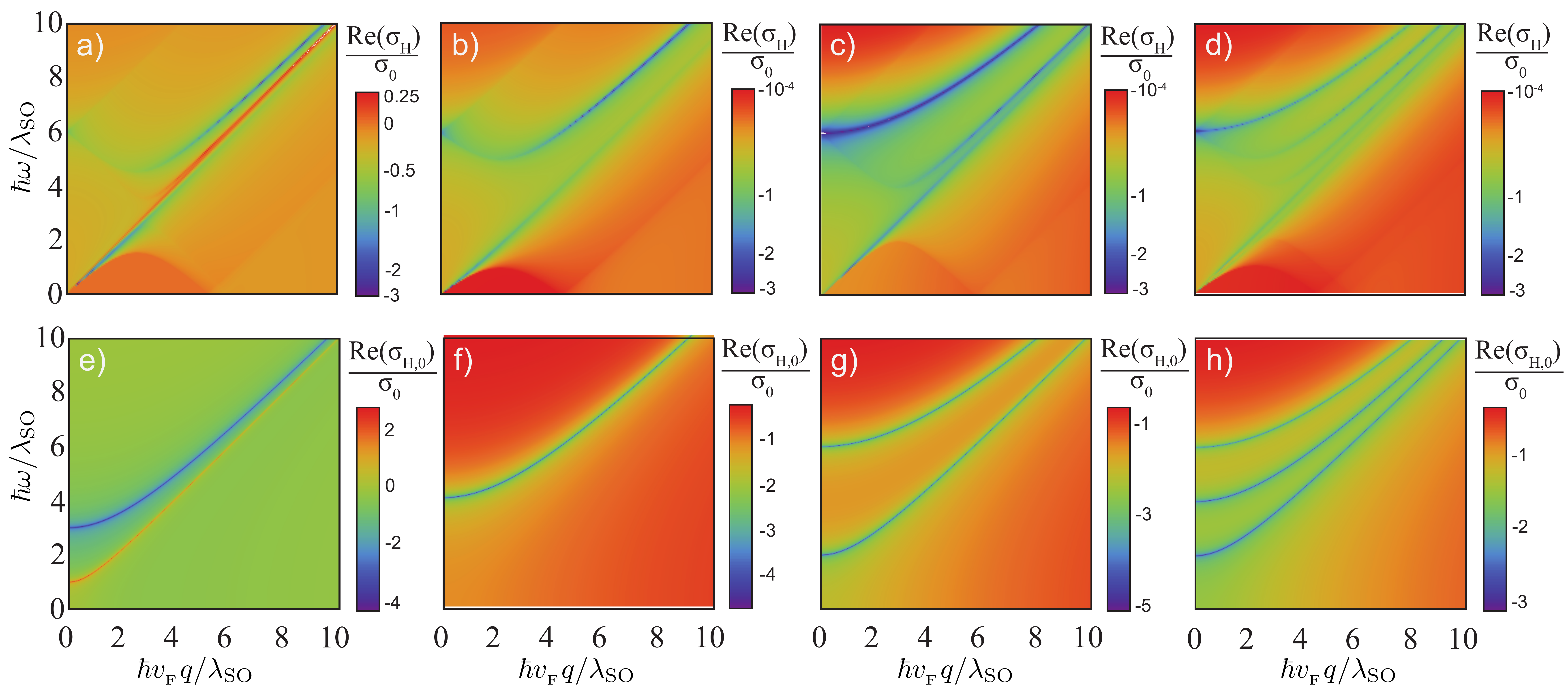}
\caption{(Color online) 
Density plots of the real part of the Hall dynamical conductivity $\sigma_H=\sigma_{H,0}+\theta(|\mu|-|\Delta|) \sigma_{H,1}$ (top panels) and its $\sigma_{H,0}$ topological contribution (bottom panels) in the $(\hbar\omega/\lambda_{SO}$, $\hbar v_F q/\lambda_{SO})$ plane for $(e \ell E_z/\lambda_{SO},\Lambda/\lambda_{SO})$ equal to $(0,0.5)$ (a,e), $(0,1)$ (b,f), $(0,2)$ (c,g), and $(0.5,1.5)$ (d,h). Parameters are the same as in Fig. 2. For all conductivities the sum over valley and spin indexes has been performed.}
\label{Fig3}
\end{figure*}

Information about the various Hall phases is contained in the nonlocal Hall conductivity $\sigma_{H}(\omega,q)$ due to its additional dependency on the sign of the gaps through the combination
$\eta \Delta^{\eta}_s$. A first consequence is that, after summing over valley and spin indexes, 
$\sigma_{H}(\omega,q)$ identically vanishes along the $\Lambda=0$ line, the contribution of spin down (up) in valley $\eta=+1$ cancels that of spin up (down) in valley $\eta=-1$. Also, because of the particular structure of $\sigma_{H}(\omega,q)$, the reflection symmetry described in the previous paragraph does not hold for the Hall conductivity. Indeed, along the $E_z=0$ line the real part of the Hall conductivity is non-vanishing (Fig. 3a-c). Similarly to the case of Fig. 2, the results of Fig. 3  can be interpreted in terms of gapped and ungapped graphene Hall conductivities. 
In Figs. 3a (QSHI phase) and 3c (AQHI phase) all gaps are opened with degeneracy equal to 2, resulting in two distinct regions in the $\omega> v_F q$ zone. Panel b) of Fig. 3 corresponds to the SPM phase, in which two gaps are opened and degenerate, and the other two gaps are closed. The latter result in a contribution to $\sigma_H$ from ungapped graphene which, of course,  is identically zero in the whole $(\omega, q)$ plane. Finally, in panel d) of Fig. 3 three opened and non-degenerate gaps contribute to the Hall conductivity.  As follows from Figs. 2 and 3, in all cases ${\rm Re} \; \sigma_H$ is orders of magnitude smaller than ${\rm Re} \; \sigma_{L/T}$, except in particular regions of the $(\omega,q)$ plane, most notably for  $ v_F q \lesssim \omega \lesssim 2 \mu - v_F q$.

Signatures of topology in the nonlocal Hall conductivity  are contained in its first term,
$\sigma_{H,0}(\omega,q)$, which can be probed when the chemical potential is smaller than all gaps (in particular, for $\mu=0$). The bottom row of Fig. 3 shows ${\rm Re} \; \sigma_{H,0}(\omega,q)$ for the same phases as in the top panels.
In the limit $\hbar \omega/\lambda_{SO},\hbar v_F q/\lambda_{SO} \ll 1$, this  component of the Hall conductivity is proportional to the Chern number, ${\rm Re} \; \sigma_{H,0} \approx (2/\pi) \sigma_0 C$; hence, for small spatial and temporal dispersion, the topological aspects of the Hall conductivity are robust. For large spatial dispersion, $v_F q \gg \omega,\Delta/\hbar$, the Hall conductivity decays as $1/q$. Noting that $\sigma_{H,0}(\omega,q)$ can be re-written as
$\frac{\sigma_0}{\pi}  \frac{\eta \Delta}{\sqrt{\Omega^2-Q^2}} \log \left[ \frac{2 |\Delta| +  \sqrt{\Omega^2-Q^2}}{ 2 |\Delta| -  \sqrt{\Omega^2-Q^2}} \right]$, it follows that for lossless materials and in the $\omega>v_F q$ regime, ${\rm Re} \; \sigma_{H,0}(\omega,q)$ presents a resonant behavior for non-closed gaps when the denominator of the log function vanishes. This results in a pronounced increase of the magnitude of Hall conductivity when $\hbar \omega= [4 \Delta^2 + (\hbar v_F q)^2]^{1/2}$, as shown in panels (e-h) of Fig. 3. 
In order to observe the resonant behavior of ${\rm Re} \; \sigma_{H,0}(\omega, q)$ in the full electronic phase diagram depicted in Fig. 1, it is necessary to either use neutral monolayers or, for finite $\mu$, to stay sufficiently away from phase transition boundaries. 
Finally, it is worth noting that when $\mu> |\Delta|$, these features are actually cancelled by similar ones present in $\sigma_{H,1}(\omega,q)$ (not shown), and hence they are not present in the full Hall conductivity ${\rm Re} \; \sigma_{H}(\omega,q)$ (top panels of Fig. 3).


\subsection{Local Limit}

From Eqs. (\ref{General_non_local_conductivity_form}-\ref{Sigma_D_mu}), we find that in the local limit of $\textbf {q}=0$ one obtains $\sigma_{L}(\omega,0) =\sigma_{T}(\omega,0)$, which results in  $\sigma_{xx}(\omega,0)=\sigma_{yy}(\omega,0)=\sigma_{L}(\omega,0)$, and  $\sigma_{xy}(\omega,0)=-\sigma_{yx}(\omega,0)=\sigma_H(\omega, 0)$. In this local limit, the conductivity components have simple analytical expressions, namely \begin{eqnarray} 
\sigma_{xx}(\omega,0) & = & i\frac{\sigma_0}{\pi} 
\left[ \frac{\mu^{2} - \Delta^{2}}{\abs{\mu}}\frac{1}{\Omega}\Theta\left[\abs{\mu} - \abs{\Delta}\right] \right. \cr
&+&  \left.\frac{\Delta^{2}}{M\Omega} - \frac{\Omega^2+4\Delta^2}{2 i \Omega^2}   \tan^{-1}\left(\frac{i\Omega}{2M}\right) \right], \cr
\sigma_{xy} (\omega,0) &=&  \frac{2 \sigma_0}{\pi}\frac{\eta\Delta}{i\Omega}\tan^{-1}\left(\frac{i\Omega}{2M}\right),
\end{eqnarray}
where $M = \text{Max}\left[\abs{\Delta}, \abs{\mu}\right]$. These results are per Dirac cone and they are consistent with the ones found by other researchers \cite{Tabert2013a,Xiao2013,WangKong2010,WangKong2011}. The first term in $\sigma_{xx}$ corresponds to intra-band transitions, and the last two terms to inter-band transitions.


\subsection{Static Limit}

\begin{figure}
\centering
\includegraphics[width=\linewidth]{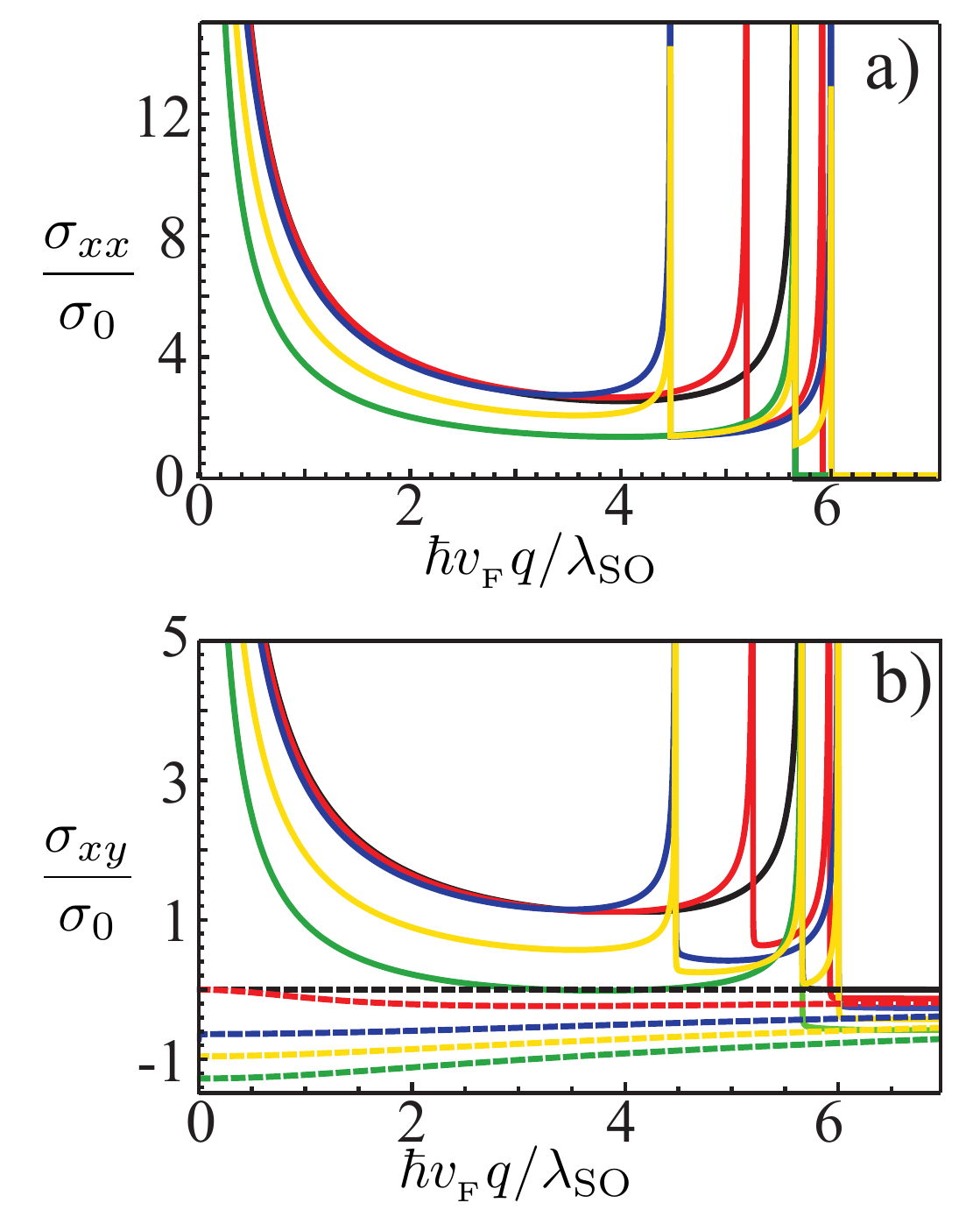}
\caption{(Color online) Real part of the static conductivities (a) $\sigma_{xx}(0,\textbf q)$ and (b) $\sigma_{xy}(0,\textbf q)$ for an arbitrarily chosen direction for the wave vector, $\varphi=\pi/3$. Their behavior for the same electronic phases as in Fig. 2 are shown: $(e \ell E_z/\lambda_{SO},\Lambda/\lambda_{SO})$ equal to $(0,0)$ (black), $(0,0.5)$ (red), $(0,1)$ (blue), $(0,2)$ (green), and $(0.5,1.5)$ (yellow). The dashed curves in panel (b) correspond to the topological part of the Hall conductivity, given by the last term in 
Eq.(\ref{Static_Conductivitiesxy}). Parameters are the same as in Fig. 2.} 
\label{Fig4}
\end{figure}

In addition to the local limit, simple analytical expressions for the conductivity at $\omega=0$ and finite wave vector $\textbf{q}$ can also be found. This static limit is directly related to the screening properties of the layered material of a charged impurity. The calculations show that the conductivity components per Dirac cone in the lossless limit $\Gamma=0$ are
\begin{eqnarray}
\label{Static_Conductivities}
\sigma_{xx}(0,\textbf{q}) & = & \frac{4 \sigma_0}{\pi}  \sin^{2}(\varphi) \frac{\mu^{2} - \Delta^{2}}{Q\sqrt{4(\mu^{2} - \Delta^{2}) - Q^{2}}}
 \nonumber\\
& & \times \Theta(2\sqrt{\mu^{2} - \Delta^{2}} -Q)\Theta( \abs{\mu} - \abs{\Delta} ) , \\
\sigma_{xy} (0,\textbf{q}) & = &  
\frac{2 \sigma_0}{\pi}\sin(2\varphi)\frac{\mu^{2} - \Delta^{2} }{Q\sqrt{4(\mu^{2} - \Delta^{2} ) - Q^{2} }} \nonumber\\
& & \times \Theta(2\sqrt{\mu^{2} - \Delta^{2}} -Q)\Theta( \abs{\mu} - \abs{\Delta} ) \cr
&+&
\frac{2\sigma_0}{\pi}\frac{\eta\Delta}{Q}\left[
\cos^{-1}\left(\frac{2|\mu|}{\sqrt{Q^{2} + 4\Delta^{2}}}\right)\Theta(|\mu|-|\Delta|) \right. \nonumber \\
& + & \left.
\tan^{-1}\left(\frac{Q}{2|\Delta|}\right)\Theta(|\Delta|-|\mu|)
\right] .
\label{Static_Conductivitiesxy}
\end{eqnarray}
Here, we have written the wave vector as $\textbf{q}=q (\cos \varphi, \sin \varphi)$. We note that $\sigma_{xx}(0,\textbf{q})$ and $\sigma_{yy}(0,\textbf{q})$ (found by substituting $\sin^2(\varphi)$ with $\cos^2(\varphi)$) are determined entirely by intraband transitions and they are nonzero when $\mu$ is greater than the mass gap. In contrast, $\sigma_{xy}(0,\textbf{q})$ and 
$\sigma_{yx}(0,\textbf{q})$ (found by flipping the sign of the second term in 
Eq.(\ref{Static_Conductivitiesxy}))
have both inter- and intraband contributions, the former contributing for $\mu<|\Delta|$. 
The $\sigma_{xx,yy}(0,\textbf{q})$ components are characterized by energy conserving transition processes which are present when $Q<2\sqrt{\mu^2-\Delta^2}$, and the same processes appear in the first term of $\sigma_{xy}(0,\textbf{q})$. 
 The disappearance of the static diagonal conductivity at $Q=2\sqrt{\mu^2-\Delta^2}$ is indicative of the absence of backscattering. Taking the limit of $\Delta=0$, we recover the static conductivity of graphene, which becomes zero when $Q=2|\mu|$ as found by others \cite{Ando2006,Pyatkovskiy2009}. 

In Fig. \ref{Fig4} we show how the real parts of $\sigma_{xx}(0,\textbf{q})$ and $\sigma_{xy}(0,\textbf{q})$ evolve as a function of $q=|\textbf{q}|$ for an arbitrarily chosen direction of the wave vector. We analyze the same electronic phases as in Fig. 2. There is a common behavior for all studied cases in $\sigma_{xx}(0,\textbf{q})$ and $\sigma_{xy}(0,\textbf{q})$, namely a Drude-like divergence for small $q$, and sharp peaks when $\hbar v_F q = 2 \sqrt{\mu^2-\Delta^2}$.  
Both of these features arise from the transverse $\sigma_T(0,q)$ contributions to the conductivity tensor, the longitudinal contributions being negligible. The dashed curves in panel b) correspond to the topological (last) term in Eq.(\ref{Static_Conductivitiesxy}) which arises from $\sigma_{H,0}$, and in the regime of weak spatial dispersion we obtain $\sigma_{H,0}(\omega=0,q \rightarrow 0)/\sigma_0= (2/\pi) C$ for each of the corresponding phases. This, in turn, implies that $\sigma_{xy}(0,0)$ is also proportional to the Chern number.


\subsection{Anisotropy due to spatial dispersion}

\begin{figure*}[t]
\includegraphics[width=1\linewidth]{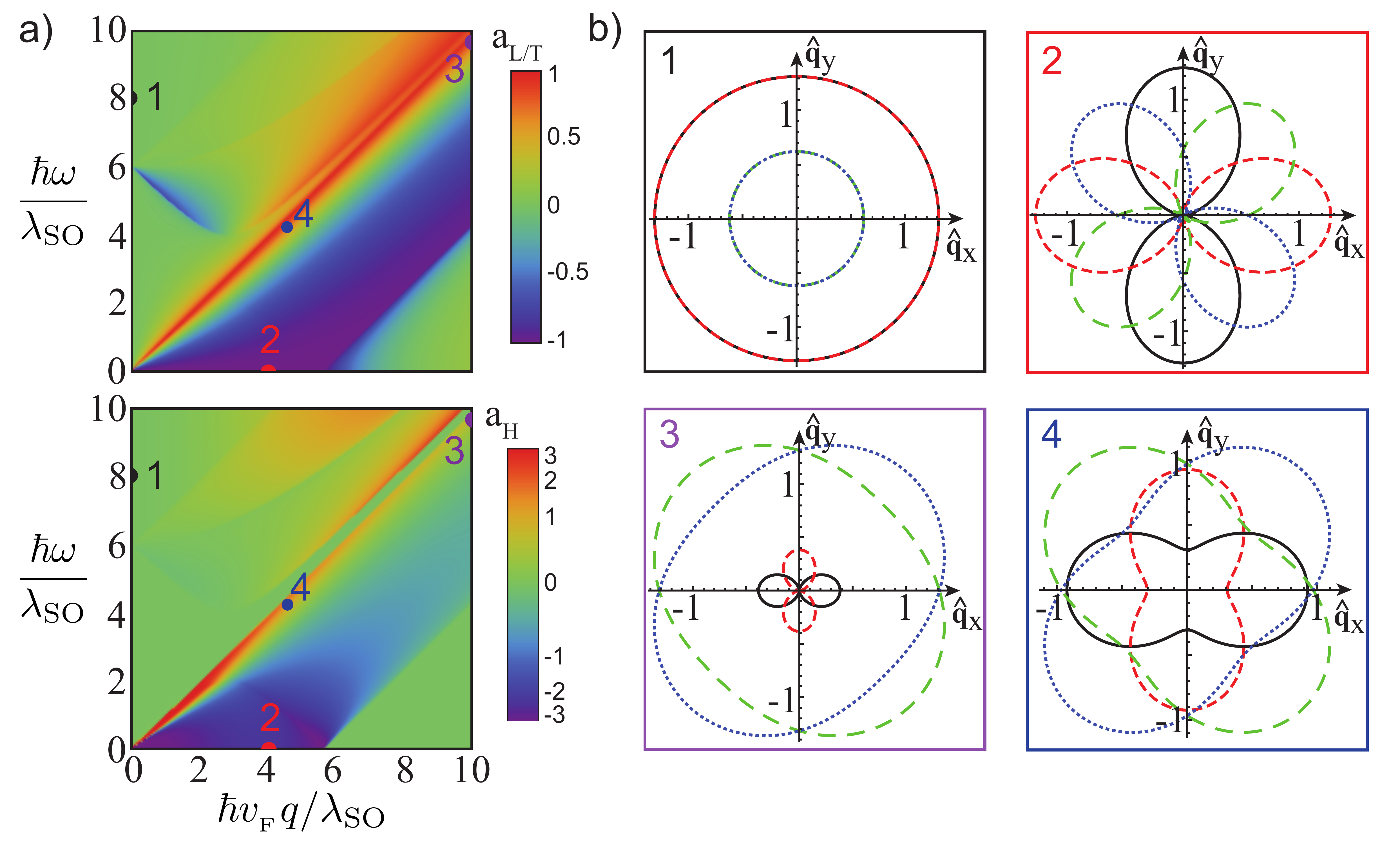}
\caption{(Color online) 
(a) Density plots of the anisotropy parameters $a_{L/T}$ and $a_H$ in the $(\omega,q)$
 plane for the graphene family materials at $(e \ell E_z/\lambda_{SO},\Lambda/\lambda_{SO})=(0,2)$ (AQHI phase).
(b) Polar plots of ${\rm Re} \sigma_{xx}/\sigma_0$ (black solid),  ${\rm Re} \sigma_{yy}/\sigma_0$ (red dashed), 
$| {\rm Re} \sigma_{xy}|/\sigma_0$ (green long dashed), and $| {\rm Re} \sigma_{yx}|/\sigma_0$ (blue dotted) for various points in the density plots of panel a): $(\hbar\omega/\lambda_{SO}$, $\hbar v_F q/\lambda_{SO})$ equal to $(8,0)$ (point 1), $(0,4)$ (point 2), $(9.86,10)$ (point 3), and $(4.42,5)$ (point 4).
Parameters are the same as in Fig. 2.
}
\label{anisotropy-parameters}
\end{figure*}

Spatial dispersion  introduces anisotropic effects in the optical response of the graphene materials with respect to the orientation of the momentum vector
$\textbf{q}$,
as evident from the $q_{i}q_{j}/q^{2}$ prefactors in Eq.~\eqref{General_non_local_conductivity_form}. As before, writing the wave vector 
$\textbf q = q (\cos \varphi, \sin \varphi)$, we
can subsequently express the nonlocal longitudinal conducitivities as $\sigma_{xx}(\omega,\textbf q)=\sigma_{T}(\omega,q)+\cos^{2}(\varphi) [\sigma_{L}(\omega, q)-\sigma_{T}(\omega,q)]$ and $\sigma_{yy}(\omega, \textbf q)$ with $\cos^{2}(\varphi)$ replaced with $\sin^{2}(\varphi)$. For the Hall conductivities, 
$\sigma_{xy,yx}(\omega,\textbf q)= \pm \sigma_{H}(\omega,q)+ \sin(\varphi) \cos(\varphi) [\sigma_{L}(\omega, q)-\sigma_{T}(\omega,q)]$. To quantify how these anisotropic effects affect the optical response, we define corresponding 
``anisotropy parameters"
\begin{eqnarray}
a_{L/T}(\omega, q) &\equiv& \frac{{\rm Re}[\sigma_{L}(\omega,q) - \sigma_{T}(\omega,q)]}{ {\rm Re}[\sigma_{L}(\omega,q) +  \sigma_{T}(\omega,q)] } , 
\label{aLT} \\
a_{H}(\omega, q) &\equiv& \frac{{\rm Re}[\sigma_{L}(\omega,q) - \sigma_{T}(\omega,q)]}{ | {\rm Re} \sigma_{H}(\omega,q)| }.
\label{aH}
\end{eqnarray}
Isotropy occurs when these anisotropy parameters vanish at a given $(\omega, q)$ point and, hence, there is no angle dependency for the corresponding conductivities. When $a_{L/T}=\pm 1$, the anisotropic effects in the diagonal conductivities are maximum, such that $\sigma_{xx}=\cos^2(\varphi) \sigma_L$ and $\sigma_{xx}=\sin^2(\varphi) \sigma_T$, respectively.
As defined, the anisotropy parameter $a_H$ measures the ratio of the anisotropic to isotropic components of the Hall conductivities $\sigma_{xy,yx}$, and ranges from $-\infty$ to $+\infty$. 
Note that Eq.(\ref{aH}) is ill-defined when $\sigma_H$ is identically zero; in this case,  a better normalization for Eq.(\ref{aH}) could be $\sigma_0$.  We show in Fig. \ref{anisotropy-parameters}a) the density plots of the anisotropy parameters for the graphene family materials at the point $(e \ell E_z/\lambda_{SO},\Lambda/\lambda_{SO})=(0,2)$ in the electronic phase space. The corresponding density plots for other phases can be easily obtained using the information provided in Figs. 2 and 3. 
Panel b) of Fig. 5 shows polar plots of the diagonal and Hall conductivities for particular points in the 
$(\omega,q)$ plane of panel a). For example, any point along the $q=0$ line in Fig. 5a), such as the chosen point 1, has $a_{L/T}=a_H=0$, and hence the corresponding polar plots for the longitudinal and Hall conductivities are circles (full isotropy). Points along the $\omega=0$ line have 
$\sigma_L \approx 0$ (see Fig. 2), and for $\hbar v_F q \lesssim 2 \mu$ we get $a_{L/T} = -1$ (maximal anisotropy for the diagonal conductivities), e.g. point 2. The values of $a_H$ along the $\omega=0$ line depends on the particular value of $q$; for example, for the chosen point 2, $a_H= -2.27$ and the corresponding polar plots for the Hall conductivities also show large anisotropy. For point 3, 
$a_{L/T} = +0.96$ (close to maximal anisotropy); note that $\sigma_{xx}$ and $\sigma_{yy}$ are rotated 90 degrees with respect to the polar plots for point 2. At point 3, $a_H=+0.28$, and the Hall conductivities are slightly anisotropic. Finally, for point 4, $a_{L/T} = +0.5$ and
$a_H= +0.64$, indicating moderate anisotropy for both the diagonal and non-diagonal conductivities.


\subsection{Plasmons in the graphene family}

\begin{figure*}[t]
\centering
\includegraphics[width=0.9\linewidth]{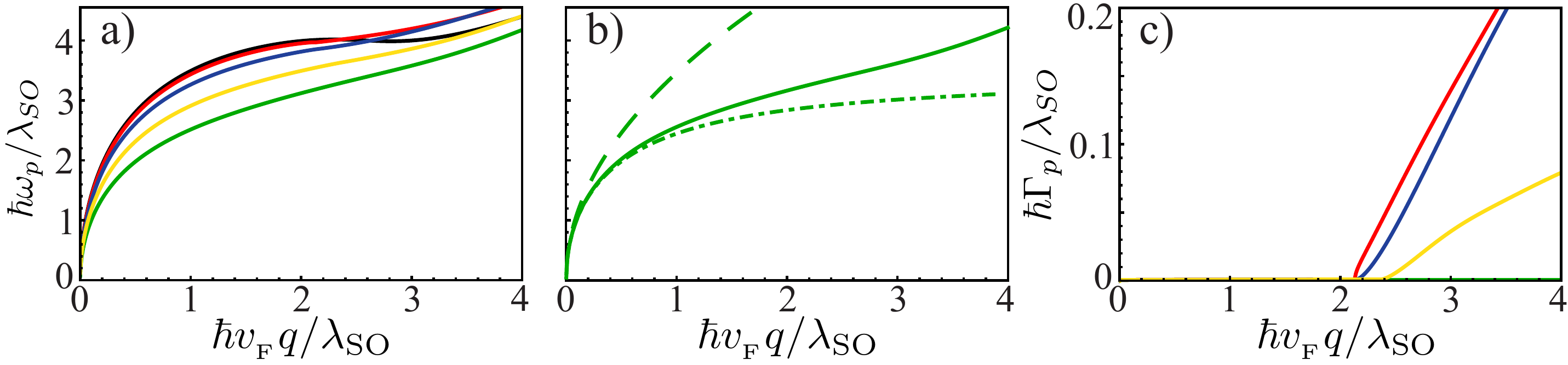}
\caption{(Color online) (a) Plasmon dispersion relation for $(e \ell E_z/\lambda_{SO},\Lambda/\lambda_{SO})$ equal to $(0,0)$ (black), $(0,0.5)$ (red), $(0,1)$ (blue), $(0,2)$ (green), and $(0.5,1.5)$ (yellow); (b) $\omega_p(q)$ vs $q$ for the $(0,2)$ point in phase space, calculated using 
Eq.(\ref{fulleq}) (solid), Eq. (\ref{Local_Plasmon_Approx_IMPROVED}) (dash-dotted), and 
$\omega_p(q)=\sqrt{\alpha c \omega_P q}$ (dashed); (c) Plasmon decay rate $\Gamma_p$
vs $q$ for the same points in phase space as in (a). Note that the curves for $(0,0)$ and $(0,2)$ are on top of each other and are negligible small in the shown $q$ range.
Other parameters are the same as in Fig. 2.
}
\label{Fig:Plasmon}
\end{figure*}

The optical response of the graphene family materials determines the electromagnetic modes they can support. Of particular interest are the plasmon surface waves, collective electronic excitations due to the coupling with an electromagnetic field. Graphene plasmons are attractive for many applications due to their strong localization, low losses, and tunability via doping \cite{Low2014,deAbajo2014,Yao2013}. The different Hall phases in the expanded graphene family accessible with external fields promise additional control of plasmon excitations. 

Based on the expressions for the conductivity tensor in Eqs. (5-10), we are in a position to investigate plasmons in this class of materials in a unified approach. Our considerations rely on the dispersion relation of the electromagnetic modes, which can be obtained by applying standard boundary conditions in a 2D layer \cite{Rodriguez-Lopez2017}. As shown in the Appendix, the poles of the Fresnel reflection matrix yield an equation for the plasmon dispersion 
\begin{eqnarray}\label{Plasmon_Numeric}
F(\omega,q) &=&\left(2\pi  q_{z}\sigma_{L}(\omega, q)+c\kappa\right)\left(2\pi \kappa  \sigma_{T}(\omega, q)+cq_{z}\right)\nonumber\\
& & + 4 \pi^{2}\kappa q_{z}\sigma_{H}^{2}(\omega, q)= 0,
\end{eqnarray}
where $\kappa = \omega/c$ and $q_{z} = \sqrt{ \kappa^{2} - q^{2} }$. One notes that 
while nonlocal effects in the optical response are captured in Eq. (\ref{Plasmon_Numeric}), anisotropy in the response does not play a role here. 
In general, the solution to Eq. (\ref{Plasmon_Numeric}) is complex and holds information about the plasmon dispersion relation and its damping. To calculate these properties, we assume $q$ is real and the $\omega$ solutions are complex, given as $\omega_p(q)  -i \Gamma_p(q)$, where $\omega_p(q)$ is the plasmon dispersion and $\Gamma_p(q)$ is the plasmon damping. Assuming $\Gamma_p(q)$ to be small, the dispersion relation and damping are found by numerically solving 
\begin{eqnarray}
&&\Imag {F(\omega_p(q), q)} \approx 0, \nonumber \\
&&\Gamma_p(q) \approx \Real{-i F(\omega_p(q),q) /  F'(\omega_p(q),q)},
\label{fulleq}
\end{eqnarray}
where the prime denotes derivative with respect to frequency. 

Significant simplifications in $\omega_p(q)$ can be achieved in the near-field approximation, where $q_z \approx i q$. By further utilizing the local approximation in $\sigma_{ij}$ (see Section III.A), and neglecting the smaller in magnitude Hall conductivity $\sigma_{H}$, one can find a simple expression for the plasmon dispersion relation
\begin{eqnarray}
\label{Local_Plasmon_Approx_IMPROVED}
q & = & \Real{\frac{i\,\omega_p(q)}{2\pi\left( \sigma_{L}^{\text{intra}}(\omega_p(q), 0) + \sigma_{L}^{\text{inter}}(\omega_p(q), 0) \right) }},
\end{eqnarray}
where we have decomposed the longitudinal conductivity in terms of its intraband $\sigma_{L}^{\text{intra}}(\omega, 0)$ and interband $\sigma_{L}^{\text{inter}}(\omega, 0)$ contributions. 
For small frequencies, the optical response is determined by the intra-band contribution only, which leads to a further simplified plasmon dispersion  $\omega_p(q) \approx \sqrt{\alpha c\omega_{P}q}$, where $\omega_{P}  =  \frac{1}{2\hbar}\sum_{\eta,s=\pm 1}\frac{\left[ \mu^{2} - (\Delta_{s}^{\mu})^{2} \right]}{\abs{\mu}}\Theta\left( \abs{\mu} - \abs{\Delta_{s}^{\eta}} \right)$ \cite{Wunsch2006,Hwang2007,Tabert2014,Wu2016}. Let us note that the $\omega_p(q) \sim \sqrt{q}$ behavior is typical for 2D systems, and that the presence of $\hbar$ in $\omega_p(q)$ indicates that quantum mechanics plays a key role in the plasmon dispersion relation of Dirac-like graphene family materials \cite{Sachdeva2015}. 

The dispersion relations given by Eqs.(\ref{fulleq}) for plasmons in distinct electronic phases are shown in Fig. \ref{Fig:Plasmon}(a). For $\hbar v_F q / \lambda_{SO} \ll 1$ all curves feature the $\sqrt{q}$ behavior mentioned above. For values of $\hbar v_F q / \lambda_{SO}$ up to around 2-3, $\omega_p(q)$ for phases with trivial topology (black and red curves) are essentially indistinguishable, while phases with non-trivial topology (blue, green, and yellow curves) possess different plasmon dispersion relations (increasing the absolute value of the Chern number $|C|$, decreases the corresponding $\omega_p(q)$). For larger values of $q$, there is no longer a clear effect of topology on the plasmons, probably due to the increased damping (see below). In Fig. \ref{Fig:Plasmon}(b) we compare the plasmon dispersion relation 
for the AQHI phase obtained from Eqs.(\ref{fulleq}) with those obtained in the local approximation. The local results practically coincide with the non-local one for small values of $q$, but as $q$ becomes larger  the plasmon dispersions obtained via the local approximations begin to depart from the non-local $\omega_p(q)$. 
The local result without inter-band contributions, 
$\omega_p(q) \approx \sqrt{\alpha c\omega_{P}q}$, overestimates the full
non-local result for $\omega_p(q)$, while Eq. (\ref{Local_Plasmon_Approx_IMPROVED}) 
underestimates it (although it provides a better approximation).
Similar trends are observed (but not shown) for other phases of these materials. 
Finally, we examine how the plasmon damping depends on the wave vector. The results in Fig. \ref{Fig:Plasmon}(c) show that for small wave vectors $\Gamma_p$ is negligible. As $q$ increases, however, the plasmon damping of the  $(e \ell E_z/\lambda_{SO},\Lambda/\lambda_{SO})=(0,0.5), (0,1)$, and $(0.5,1.5)$ points in phase space increases, while it remains small for the other two phases (indicating that the plasmons are well defined modes in the considered $q-$range for these two phases). It appears that spatial dispersion in the optical response of the graphene materials electronic phases can lead to a variety of behaviors of their plasmon characteristics. 


\section{Conclusions}

In this paper, a unified description of the dispersive non-local optical response of the graphene family materials is presented, capturing the different Hall phases of these materials accessible via an external electrostatic field and circularly polarized laser light. The explicit expressions for all components of the conductivity tensor are useful to understand the interplay between effects of frequency and spatial dispersion and Hall phase transitions in these systems. 
A comprehensive understanding of the optical anisotropy in these materials induced by spatial nonlocality is also presented. Our calculations also show that spatial dispersion can affect plasmonic properties in the graphene family, especially at large wave vectors. Altogether, our 
work provides a full description of the nonlocal optical response in topological phase transitions present in 2D staggered semiconductors, and similar analysis could be used in emergent van der Waals materials. 


\section*{Acknowledgements}
We acknowledge financial support from the US Department of Energy under grant No. DE-FG02-06ER46297, the LANL LDRD program, and CNLS. P.R.-L. also acknowledges partial support from TerMic (Grant No. FIS2014-52486-R, Spanish Government) and from Juan de la Cierva - Incorporacion program.

%


\begin{widetext}
\appendix
\section{Calculation Procedure of the Conductivity Tensor Components}\label{Velocity Operator}
\renewcommand{\theequation}{A.\arabic{equation}}
\setcounter{equation}{0}

The decomposition of the optical conductivity into longitudinal, transverse, and Hall components, as shown in Eq. 4, can be obtained by considering the matrix elements of the velocity operators entering Eq. 3. Using the definition 
 $\mean{v_{i}v_{j}}_{\textbf{k},\textbf{k}+\textbf{q}}^{\lambda,\lambda'} =  \bra{u_{\textbf{k}}^{\lambda}} v_{i} \ket{u_{\textbf{k}+\textbf{q}}^{\lambda'}} \bra{u_{\textbf{k}+\textbf{q}}^{\lambda'}} v_{j} \ket{u_{\textbf{k}}^{\lambda}}$, we find 
\begin{eqnarray}
\mean{v_{x}v_{x}}_{\textbf{k},\textbf{k}+\textbf{q}}^{\lambda,\lambda'} & = & \frac{v_{F}^{2}}{2}\left[ 1 + \frac{\hbar^{2}v_{F}^{2}\left[ k_{x}(k_{x} + q_{x}) - k_{y}(k_{y} + q_{y}) \right] - \Delta^{2}}{\epsilon_{\textbf{k}}^{\lambda}\epsilon_{\textbf{k}+\textbf{q}}^{\lambda'}} \right], \nonumber\\
\mean{v_{x}v_{y}}_{\textbf{k},\textbf{k}+\textbf{q}}^{\lambda,\lambda'} & = & \frac{v_{F}^{2}}{2}\left[ i\eta\Delta\left( \frac{1}{\epsilon_{\textbf{k}}^{\lambda}} - \frac{1}{\epsilon_{\textbf{k}+\textbf{q}}^{\lambda'}} \right) - \frac{\hbar^{2}v_{F}^{2}\left[ k_{x}(k_{y} + q_{y}) + k_{y}(k_{x} + q_{x}) \right]}{\epsilon_{\textbf{k}}^{\lambda}\epsilon_{\textbf{k}+\textbf{q}}^{\lambda'}} \right], \nonumber\\
\mean{v_{y}v_{x}}_{\textbf{k},\textbf{k}+\textbf{q}}^{\lambda,\lambda'} & = & \frac{v_{F}^{2}}{2}\left[ - i\eta\Delta\left( \frac{1}{\epsilon_{\textbf{k}}^{\lambda}} - \frac{1}{\epsilon_{\textbf{k}+\textbf{q}}^{\lambda'}} \right) - \frac{\hbar^{2}v_{F}^{2}\left[ k_{x}(k_{y} + q_{y}) + k_{y}(k_{x} + q_{x}) \right]}{\epsilon_{\textbf{k}}^{\lambda}\epsilon_{\textbf{k}+\textbf{q}}^{\lambda'}} \right], \nonumber\\
\mean{v_{y}v_{y}}_{\textbf{k},\textbf{k}+\textbf{q}}^{\lambda,\lambda'} & = & \frac{v_{F}^{2}}{2}\left[ 1 + \frac{\hbar^{2}v_{F}^{2}\left[ - k_{x}(k_{x} + q_{x}) + k_{y}(k_{y} + q_{y}) \right] - \Delta^{2}}{\epsilon_{\textbf{k}}^{\lambda}\epsilon_{\textbf{k}+\textbf{q}}^{\lambda'}} \right].
\end{eqnarray}

We further utilize polar coordinates for $\textbf{k}$ and $\textbf{q}$, such that $\textbf{k}=(k\cos(\theta+\varphi), k\sin(\theta+\varphi))$ and $\textbf{q}=(q\cos(\varphi), q\sin(\varphi))$, recasting the velocity operators in the following form,

\begin{eqnarray}\label{Conductivity_General_expression}
\mean{v_{i}v_{j}}_{\textbf{k},\textbf{k}+\textbf{q}}^{\lambda,\lambda'} & = & 
\frac{q_{i}q_{j}}{q^{2}}\mean{v^{2}}_{L}^{\lambda,\lambda'}
  +\left( \delta_{ij} - \frac{q_{i}q_{j}}{q^{2}} \right)\mean{v^{2}}_{T}^{\lambda,\lambda'}
 + \epsilon_{ij}\mean{v^{2}}_{H}^{\lambda,\lambda'}
 + S_{ij}\mean{v^{2}}_{S}^{\lambda,\lambda'},
\end{eqnarray}
\begin{eqnarray}\label{Correlaciones_velocidad_velocidad}
\mean{v^{2}}_{T}^{\lambda,\lambda'} & = & \frac{v_{F}^{2}}{2}\left[ 1 - \frac{\hbar^{2}v_{F}^{2}\left[ k\,q\cos(\theta) + k^{2}\cos(2\theta) \right] + \Delta^{2}}{\epsilon_{\textbf{k}}^{\lambda}\epsilon_{\textbf{k}+\textbf{q}}^{\lambda'}} \right],\nonumber\\
\mean{v^{2}}_{L}^{\lambda,\lambda'} & = & \frac{v_{F}^{2}}{2}\left[ 1 + \frac{\hbar^{2}v_{F}^{2}\left[ k\,q\cos(\theta) + k^{2}\cos(2\theta) \right] - \Delta^{2}}{\epsilon_{\textbf{k}}^{\lambda}\epsilon_{\textbf{k}+\textbf{q}}^{\lambda'}} \right],\nonumber\\
\mean{v^{2}}_{H}^{\lambda,\lambda'} & = & \frac{v_{F}^{2}}{2}\left[ - i\eta\Delta\left( \frac{1}{\epsilon_{\textbf{k}}^{\lambda}} - \frac{1}{\epsilon_{\textbf{k}+\textbf{q}}^{\lambda'}} \right)\right],\nonumber\\
\mean{v^{2}}_{S}^{\lambda,\lambda'} & = & \frac{v_{F}^{2}}{2}\left[ \frac{\hbar^{2}v_{F}^{2}\left[ k\,q\sin(\theta) + k^{2}\sin(2\theta) \right]}{\epsilon_{\textbf{k}}^{\lambda}\epsilon_{\textbf{k}+\textbf{q}}^{\lambda'}} \right],\nonumber\\
S_{ij} & = & \left(\begin{array}{cc}
- \sin(2\varphi) & \cos(2\varphi)\\
  \cos(2\varphi) & \sin(2\varphi)
\end{array}\right),
\end{eqnarray}
where $\epsilon_{ij}$ is the 2D Levi-Civita symbol.  Clearly, the first three terms in Eq. \ref{Conductivity_General_expression} correspond to $\sigma_{L}$, $\sigma_{T}$, and $\sigma_{H}$, respectively. The last term, however, is an odd function of the  angular variable $\theta$ in $\mean{v^{2}}_{S}^{\lambda,\lambda'}$, thus it makes no contribution after integration in the Kubo formula and it is not considered further. 

Using the above results, we also show how $\sigma_{p}(\omega, q$), $p=\{L, T, H\}$ is represented by distinguishing the role of the chemical potential as given as $\sigma_{p}(\omega, q) =  \sigma_{p,0}(\omega, q) + \Theta(\abs{\mu} - \abs{\Delta})\sigma_{p,1}(\omega, q)$. After substituting $\mean{v^{2}}_{p}^{\lambda,\lambda'}$ into Eq. \ref{General_Kubo_Formula} of the main text, the $d^{2}\bm{k}$ integration is performed as required by the general Kubo expression. For this purpose, we use $d^2 \textbf {k} = dk_x dk_y$ in the terms containing $n_F (\epsilon_{\textbf k}^\lambda)$, while the change of variables $(k_x, k_y)\rightarrow -(k_x+q_x, k_y+q_y)$ is taken for the integration of the terms containing $n_F (\epsilon_{{\textbf k}+\textbf {q}}^\lambda)$. Taking into account that  the Fermi-Dirac distribution is $n_{F}(\epsilon_{\textbf{k}}^{\lambda})=\Theta( \mu - \epsilon_{\textbf{k}}^{\lambda} )$ at $T=0$ allows to separate  the $\Theta( \mu + \epsilon_{\textbf{k}}^{\lambda})=1$ from the $\Theta( \mu - \epsilon_{\textbf{k}}^{\lambda})$ contributions. Thus one arrives at
\begin{eqnarray}\label{Integral_mu_cero}
\sigma_{p,0} (\omega, q) =  i\frac{\alpha c\hbar^{2}}{(2\pi)^{2}}\int dk_{x}\int dk_{y}\Theta( \epsilon_{c}^{2} - \epsilon_{k}^{2} )
\mean{v^{2}}_{p}^{+,-}\left( \frac{1}{\Omega - \epsilon_{k} - \epsilon_{kq}}\frac{1}{ \epsilon_{k} +\epsilon_{kq}} + \frac{(1-2\delta_{p,H})}{\Omega + \epsilon_{k} + \epsilon_{kq}}\frac{1}{ \epsilon_{k} + \epsilon_{kq}} \right),
\end{eqnarray}
\begin{eqnarray}\label{Integral_mu_NO_cero}
\sigma_{p,1} (\omega, q) = i\frac{\alpha c\hbar^{2}}{(2\pi)^{2}}\int dk_{x}\int dk_{y}\Theta( \mu^{2} - \epsilon_{k}^{2} )
&&
\left[ 
-
\mean{v^{2}}_{p}^{+,-}\left( \frac{1}{\Omega - \epsilon_{k} - \epsilon_{kq}}\frac{1}{ \epsilon_{k} + \epsilon_{kq}} + \frac{(1-2\delta_{p,H})}{\Omega + \epsilon_{k} + \epsilon_{kq}}\frac{1}{ \epsilon_{k} + \epsilon_{kq}} \right) \right. \cr
&&  \left. + \mean{v^{2}}_{p}^{+,+}\left( \frac{1}{\Omega - \epsilon_{k} + \epsilon_{kq}}\frac{1}{ - \epsilon_{k} + \epsilon_{kq}} - \frac{(1-2\delta_{p,H})}{\Omega + \epsilon_{k} - \epsilon_{kq}}\frac{1}{ \epsilon_{k} - \epsilon_{kq}} \right)
\right],
\end{eqnarray}
where $\epsilon_{k}=\sqrt{\hbar^{2} v_{F}^{2}k^{2} + \Delta^{2}}$, $\epsilon_{kq}=\sqrt{\epsilon_{k}^{2} + \hbar^{2} v_{F}^{2}q^{2} + 2\hbar^{2} v_{F}^{2}kq\cos(\theta)}$, and $\delta_{p,H} = 1$ when $p=H$ and zero otherwise. The upper energy cut-off $\epsilon_{c}$ is taken to be infinity at the end of the calculations.

To obtain the results in Eqs.~\eqref{Sigma_L_0}, \eqref{Sigma_T_0} and \eqref{Sigma_D_0} the change of variables $(k_{x}, k_{y}) \rightarrow (\epsilon_{k}, B = \cos(\theta))$ is applied to Eq.~\eqref{Integral_mu_cero}. Then, by using the Sokhotski-Plemelj theorem, $\lim_{\Gamma\to 0^{+}}\frac{1}{x \pm i\Gamma}  =  \mathcal{P}\left[\frac{1}{x}\right] \mp i\pi\delta(x)$, and retaining only the Dirac delta function one calculates the $\Real{\sigma_{L,0}}$, $\Real{\sigma_{T,0}}$, and $\Imag{\sigma_{H,0}}$ for real $\omega$. By taking advantage of the Dirac delta function properties the integration over $B$ is performed. The remaining integral over $\epsilon_{k}$ is carried out using the change of variable $x=(\hbar\omega + 2\lambda\epsilon_{k})/(\hbar v_{F}q R)$, where $R=\sqrt{ 1 + 4\Delta^{2}/(Q^{2} - \Omega^{2})}$. At this point, instead of using the Kramers-Kr\"onig relation to obtain the imaginary (real) part of the conductivity at real frequencies as done by several authors \cite{Pyatkovskiy2009,Tabert2014,Hwang2007,Wunsch2006,Rodriguez-Lopez2014}, we can obtain more general expressions of the different components at all complex frequencies $\Omega = \hbar(\omega + i\Gamma)$ in the upper complex plane.  Such expressions are advantageous in examining finite carrier relaxation effects. Here we apply the Kramers-Kr\"onig formula (valid for real frequencies) in order to find the real part of the conductivity at imaginary frequencies $\omega = i\xi$
\begin{eqnarray}
\Real{\sigma_{p,0}(i\xi)} & = & \frac{2}{\pi}\int_{0}^{\infty}d\omega\frac{\xi}{\omega^{2} + \xi^{2}}\Real{\sigma_{p,0}(\omega)} = \frac{2}{\pi}\int_{0}^{\infty}d\omega\frac{\omega}{\omega^{2} + \xi^{2}}\Imag{\sigma_{p,0}(\omega)}.
\end{eqnarray}
Realizing that $\Imag{\sigma_{p,0}(i\xi)} = 0$, the analytical continuation of the obtained $\Real{\sigma_{p,0}(i\xi)}$ to all positive imaginary frequencies in the upper half of the complex plane yields the final conductivity components. By taking $\xi \to\xi + \Gamma $ finite dissipation can be added to the optical response, as shown in Eqs.~\eqref{Sigma_L_0}, \eqref{Sigma_T_0} and \eqref{Sigma_D_0} in the main text.

The calculations for $\sigma_{p, 1}$ are similar. In this case, however, the integration over $B = cos(\theta)$ can be carried out directly for complex frequencies $\Omega = \hbar(\omega+i\Gamma)$. An important step here is that the contour cuts in the complex plane arising from the logarithmic functions have to be placed below the real axis to avoid the appearance of spurious cuts and changes of sign. The remaining integral over $\epsilon_{k}$ is carried out easily with the change of variable to $x$ defined above. As a result, one obtains Eqs.~\eqref{Sigma_L_mu}, \eqref{Sigma_T_mu} and \eqref{Sigma_D_mu} with the following auxiliary functions
\begin{eqnarray}
\mathcal{F}_{1} & = & 
 \phantom{-i}  (2 \abs{\Delta} - \Omega)\sqrt{ Q^{2}R^{2}-( \Omega - 2\abs{\Delta} )^{2} }
 -\phantom{i}  (2 \abs{   \mu} - \Omega)\sqrt{ Q^{2}R^{2}-( \Omega - 2\abs{   \mu} )^{2} }\nonumber\\
& &  - i(2 \abs{\Delta} + \Omega)\sqrt{ (2\abs{\Delta} +\Omega )^{2} - Q^{2}R^{2} }
 + i(2 \abs{   \mu} + \Omega)\sqrt{ (2\abs{   \mu} +\Omega )^{2} - Q^{2}R^{2} },\\
\mathcal{F}_{2} & = & 
 \phantom{-i}    \tan^{-1}\left(\frac{\Omega -2\abs{   \mu} }{\sqrt{ Q^{2}R^{2} - ( \Omega - 2\abs{   \mu} )^{2} } }\right)
 -  \tan^{-1}\left(\frac{\Omega -2\abs{\Delta} }{\sqrt{ Q^{2}R^{2} - ( \Omega - 2\abs{\Delta} )^{2} } }\right)\nonumber\\
& &  - i\log\left( \Omega + 2\abs{   \mu} +\sqrt{ ( \Omega + 2\abs{   \mu} )^{2} - Q^{2}R^{2} } \right)
 + i\log\left( \Omega + 2\abs{\Delta} +\sqrt{ ( \Omega + 2\abs{\Delta} )^{2} - Q^{2}R^{2} } \right),
\\
\mathcal{F}_{3} & = & \phantom{+} 2\mu\left( \sqrt{ Q^{2} - 4\left( \mu^{2} - \Delta^{2} \right) } + i\sqrt{ 4\left( \mu^{2} - \Delta^{2} \right) - Q^{2} }\right)\nonumber\\
& & + \left(4\Delta^{2} - Q^{2}\right)\left[\begin{array}{l}
   i\log( 2\abs{\Delta} + iQ )
 - i\log\left( 2\abs{\mu} + \sqrt{ 4\left( \mu^{2} - \Delta^{2} \right) - Q^{2} }\right)\\
\end{array}\right]\nonumber\\
& & + \left(4\Delta^{2} - Q^{2}\right)\left[\begin{array}{l}
    \tan^{-1}\left(\frac{2\abs{\mu}}{\sqrt{ Q^{2} - 4(\mu^{2} - \Delta^{2}) }}\right)
 -  \tan^{-1}\left(\frac{2\abs{\Delta}}{Q}\right)
\end{array}\right],\\
\mathcal{F}_{4} & = & - \frac{\mu}{2}\sqrt{ Q^{2} - 4\left( \mu^{2} - \Delta^{2} \right) }
 + \left( \frac{Q^{2}}{4} - \Delta^{2} \right) \left[ \tan^{-1}\left(\frac{2\abs{\mu}}{\sqrt{ Q^{2} - 4(\mu^{2} - \Delta^{2}) }}\right) - \tan^{-1}\left(\frac{2\abs{\Delta}}{Q}\right)\right],\\
\mathcal{F}_{5} & = & \left( \frac{Q^{2}}{4} - \Delta^{2} \right)\left[ \frac{\pi}{2} - \tan^{-1}\left(\frac{2\abs{\Delta}}{Q}\right)\right].
\end{eqnarray}

Let us also note that the components of the dynamical conductivity are necessary to obtain the Fresnel reflection matrix characterizing the electromagnetic boundary conditions for the considered layered system. The fact that the diagonal elements of the conductivity tensor are different is attributed to the spatial dispersion effects from the wave vector. The off-diagonal elements arise from the Hall effects in the graphene family. Imposing standard boundary conditions to a single 2D layer, one finds
\begin{equation}
R_{\text{ss}} =  - \frac{2\pi}{\delta}\left[ \frac{\sigma_{T}(\omega, q)}{c\lambda} + \frac{2\pi}{c^{2}}\Det{\sigma(\omega, q)} \right], \;
R_{\text{sp}} = R_{\text{ps}} =  \frac{2\pi \sigma_{H}(\omega, q)}{\delta c}, \;
R_{\text{pp}} =  \frac{2\pi}{\delta}\left[ \lambda\frac{\sigma_{L}(\omega, q)}{c} + \frac{2\pi}{c^{2}}\Det{\sigma(\omega, q)} \right],
\end{equation}
where
$\delta =1 + 2\pi\left( \lambda\frac{\sigma_{L}(\omega, q)}{c} + \frac{1}{\lambda}\frac{\sigma_{T}(\omega, q)}{c} \right) + \frac{4\pi^{2}}{c^{2}}\Det{\sigma(\omega, q)}$, 
$\Det{\sigma(\omega, q)} = \sigma_{L}(\omega, q)\sigma_{T}(\omega, q) + \sigma_{H}^{2}(\omega, q)$, $\lambda = q_{z}c/\omega$, and $q_z=\sqrt{\kappa^2-q^2}$. The determinant of the Fresnel matrix can further be expressed as 
\begin{eqnarray}
\Det{R} & = & \frac{ - 4\pi^{2}q_{z}\kappa \Det{\sigma(\omega, q)}}{\left(2\pi  q_{z}\sigma_{L} + c\kappa\right)
\left(2\pi \kappa  \sigma_{T} + cq_{z}\right) + 4 \pi^{2}\kappa\sigma_{H}^{2}q_{z}},
\end{eqnarray}
where $\kappa=\omega /c$. Taking that $\Det{R( \omega_{p} - i\gamma, q )}^{-1} = 0$, we obtain Eq.~\eqref{Plasmon_Numeric} in the main text.
\end{widetext}

\end{document}